\documentclass[twocolumn,twocolappendix]{aastex6}

\usepackage {epsfig,graphicx,color}
\usepackage {txfonts   }
\usepackage {natbib    }
\usepackage {float     }
\usepackage {enumerate }
\usepackage {tabularx  }
\usepackage {url       }
\usepackage {capt-of   }
\usepackage {color     }

\begin{document}

%%%%%%%%%%%%%%%%%%%%%%%%%%%%%%%%%%%%%%%%%%%%%%%%
%
% Title / Abstract%
%
%%%%%%%%%%%%%%%%%%%%%%%%%%%%%%%%%%%%%%%%%%%%%%%%

\title{Cloudless atmospheres for young low-gravity substellar objects}

\author{
P. Tremblin\altaffilmark{1}             and
G. Chabrier\altaffilmark{2,3}           and
I. Baraffe\altaffilmark{2,3}            and
Michael. C. Liu\altaffilmark{4,5}       and
E. A. Magnier\altaffilmark{4}           and
P.-O. Lagage\altaffilmark{6,7}          and
C. Alves de Oliveira\altaffilmark{8  }  and
A. J. Burgasser\altaffilmark{9  }       and
D. S. Amundsen\altaffilmark{10,11}      and 
B. Drummond\altaffilmark{2}     
       }
\altaffiltext{1}{
  Maison de la Simulation, CEA, CNRS, Univ. Paris-Sud, UVSQ, Universit\'e Paris-Saclay, 91191 Gif-sur-Yvette, France}

\altaffiltext{2}{
  Astrophysics Group, University of Exeter, EX4 4QL Exeter, UK}

\altaffiltext{3}{
  Ecole Normale Sup\'erieure de Lyon, CRAL, UMR CNRS 5574, 69364 Lyon
  Cedex 07, France}

\altaffiltext{4}{
  Institute for Astronomy, University of Hawaii, 2680 Woodlawn Drive, Honolulu, HI 96822, USA}

\altaffiltext{5}{
Visiting Astronomer at the Infrared Telescope Facility, which is operated by the University of Hawaii under Cooperative Agreement no. NNX-08AE38A with the National Aeronautics and Space Administration, Science Mission Directorate, Planetary Astronomy Program. }

\altaffiltext{6}{
  Irfu, CEA, Universit\'e Paris-Saclay, F-9119 Gif-sur Yvette, France}

\altaffiltext{7}{
Universit\'e Paris Diderot, AIM, Sorbonne Paris Cit\'e, CEA, CNRS, F-91191 Gif-sur-Yvette, France}

\altaffiltext{8}{
European Space Agency, c/o STScI, 3700 San Martin Drive, Baltimore, MD 21218, USA}

\altaffiltext{9}{
UC San Diego, M/C 0424, 9500 Gilman Drive, La Jolla, CA 92093, USA
}

\altaffiltext{10}{Department of Applied Physics and Applied
  Mathematics, Columbia University, New York, NY 10025, USA}

\altaffiltext{11}{NASA Goddard Institute for Space Studies, New York,
  NY 10025, USA} 

\email{pascal.tremblin@cea.fr}

\begin{abstract}
  Atmospheric modeling of low-gravity (VL-G) young brown dwarfs remains a challenge. The presence of very thick clouds has been suggested because of their extremely red near-infrared (NIR) spectra, but no cloud models provide a good fit to the data with a radius compatible with evolutionary models for these objects. We show that cloudless atmospheres assuming a temperature gradient reduction caused by fingering convection provides a very good model to match the observed VL-G NIR spectra. The sequence of extremely red colors in the NIR for atmospheres with effective temperature from $\sim$~2000~K down to $\sim$~1200~K is very well reproduced with predicted radii typical of young low-gravity objects. Future observations with NIRSPEC and MIRI on the James Webb Space Telescope (JWST) will provide more constrains in the mid-infrared, helping to confirm/refute whether or not the NIR reddening is caused by fingering convection. We suggest that the presence/absence of clouds will be directly determined by the silicate absorption features that can be observed with MIRI. JWST will therefore be able to better characterize  the atmosphere of these hot young brown dwarfs and their low-gravity exoplanet analogues.
\end{abstract}


\maketitle

%%%%%%%%%%%%%%%%%%%%%%%%%%%%%%%%%%%%%%%%%%%%%%%%
%
% Introduction
%
%%%%%%%%%%%%%%%%%%%%%%%%%%%%%%%%%%%%%%%%%%%%%%%%

\section{Introduction}\label{sect:intro}

Although representing a small fraction of the entire brown dwarf population, young brown dwarfs ($\leq$~150~Myrs) largely populate and even dominate the redward part (J-K$\gtrsim$ 0.8) of the color-magnitude diagram \citep[e.g.][]{chabrier:2002aa} and are of great interest because of their large radii and low surface gravity. Their study provides a promising approach to better understand low-gravity ultracool analog atmospheres such as those of young gas giant exoplanets. Recently, wide-field surveys have led to the identification of large samples of such low-gravity objects \citep[e.g.][]{Gagne:2015dc,aller:2016aa}. This population forms a separate sequence in near-infrared (NIR) color-magnitude diagrams (CMD) from the field objects \citep{Liu:2013gy,Liu:2016co,Faherty:2016fx}. This sequence is typically $\sim$~0.5 mag redder in $J-K$ color and has been interpreted as the presence of thicker clouds with small particles, even though current cloud models  struggle to fit these data \citep[e.g.][]{Liu:2016co}. The models of \citet{Saumon:2008im,Marley:2012fo} cannot reach the reddest colors at a given magnitude although they have provided reasonable fits and radii for the HR8799 planets \citep{Marley:2012fo} and BT-Settl models lead to implausibly small fitted radii \citep[e.g.][]{Liu:2013gy}.

Recently, \citet{Tremblin:2016hi} have proposed that the reddening of NIR colors in standard field L dwarfs and exoplanet analogs such as HR8799c is induced by the reduction of the temperature gradient in their atmospheres. The development of fingering convection caused by the mean molecular weight gradient implied by the chemical transition between CO and CH$_4$ could be at the origin of such a temperature gradient reduction. This mechanism naturally explains why the disappearance of the reddening at the L/T transition is concomitant with the transition between CO and CH$_4$ in the atmosphere of these objects. 

In this paper, we show that the same model based on \citet{Tremblin:2016hi} reproduce very well spectra of VL-G objects (DENIS J1425-36, 2MASS J2208+29, and PSO J318-22, chosen as late-L members of well-established young moving groups AB Doradus and Beta Pictoris) as well as the sequence in M$_J$ versus $J-K$ CMD. Furthermore, the radii inferred from the spectral fits are in good agreement with evolutionary models of young objects. The derived ages of DENIS J1425-36 and PSO J318-22 are in good agreement with their ages estimated from their membership of their respective moving groups. Therefore, these results support the CO/CH$_4$ fingering-unstable interpretation of the L/T transition of standard field objects and VL-G objects. 

\begin{figure*}[t]
\centering
\includegraphics[width=0.48\linewidth]{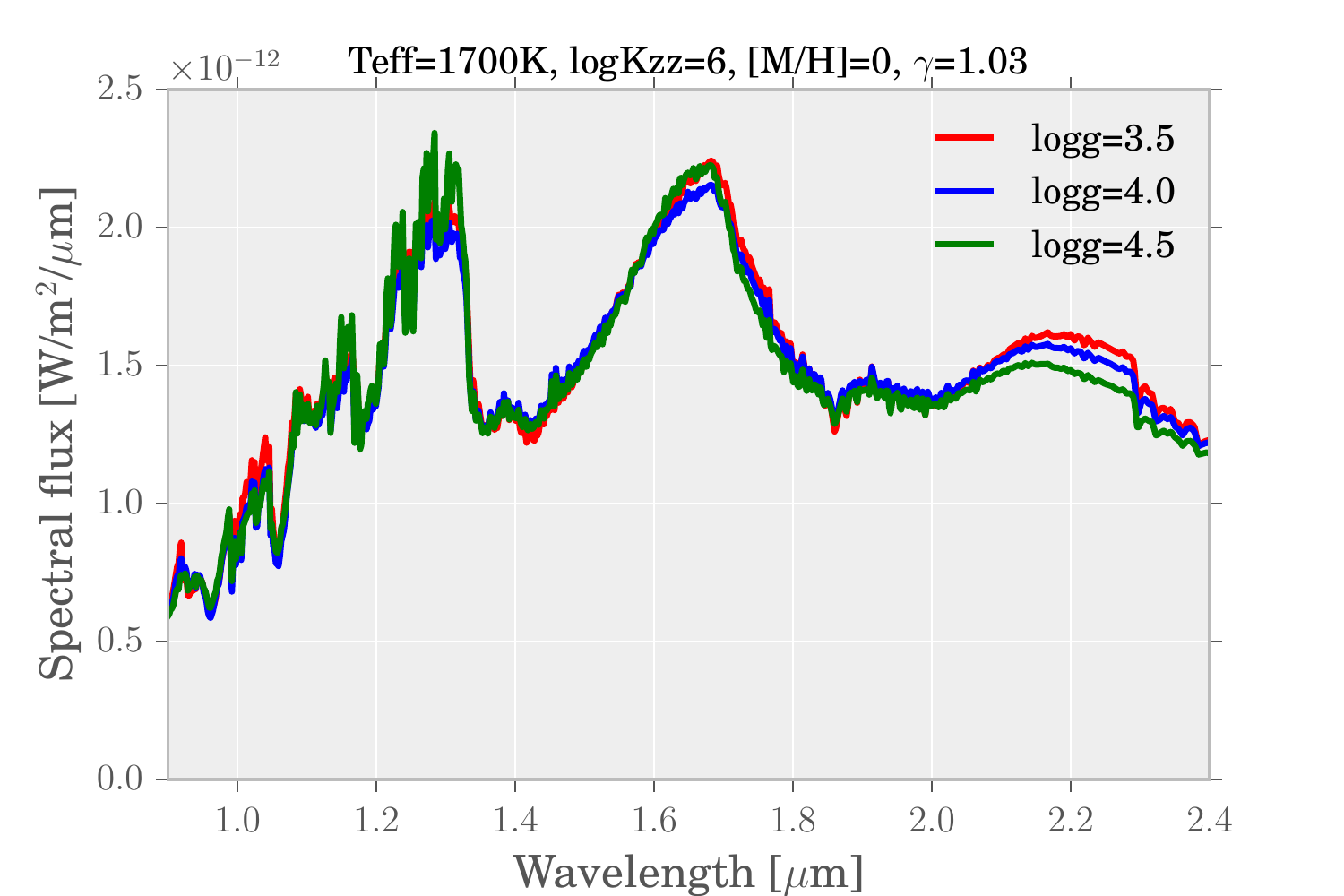}
\includegraphics[width=0.48\linewidth]{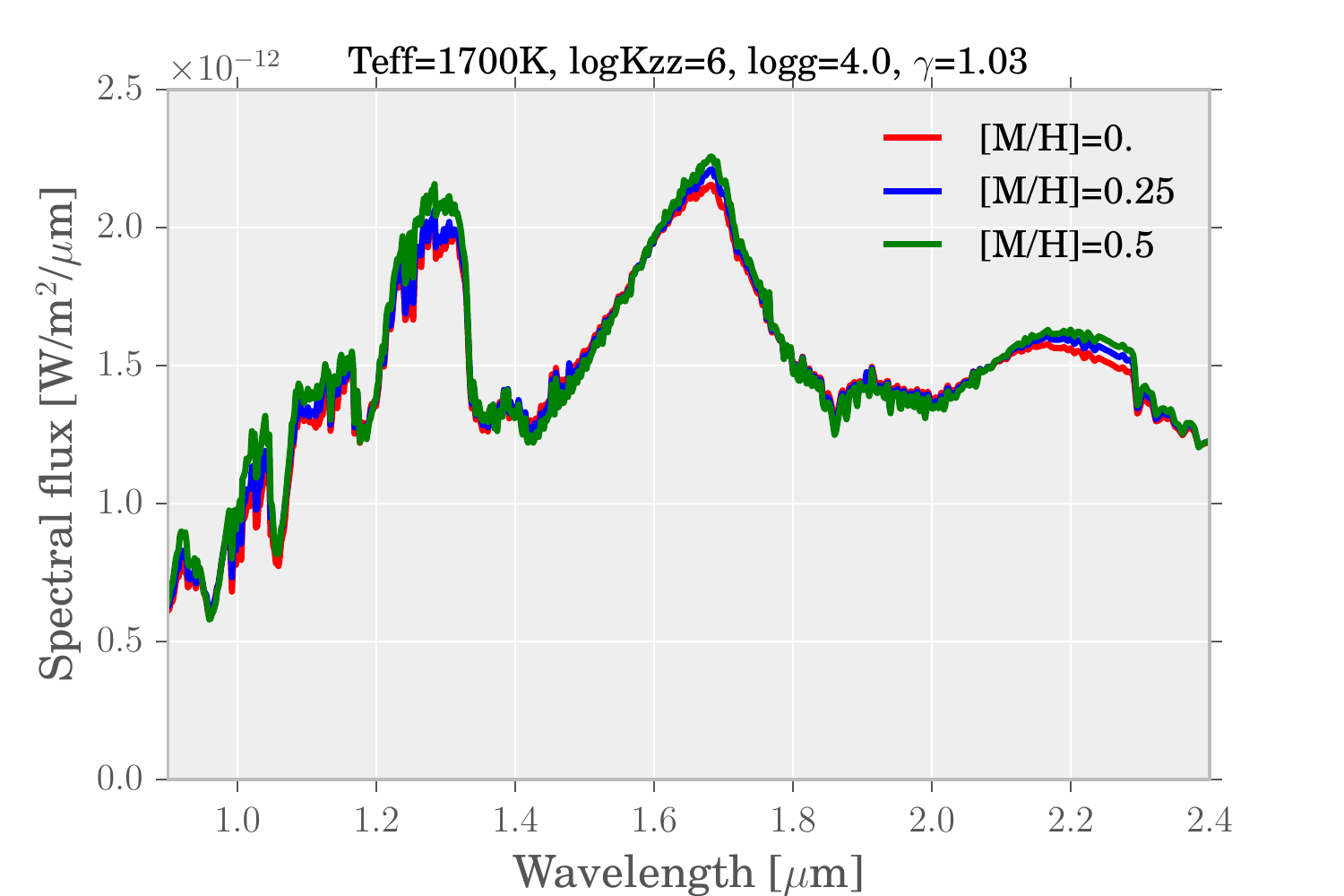}
\includegraphics[width=0.48\linewidth]{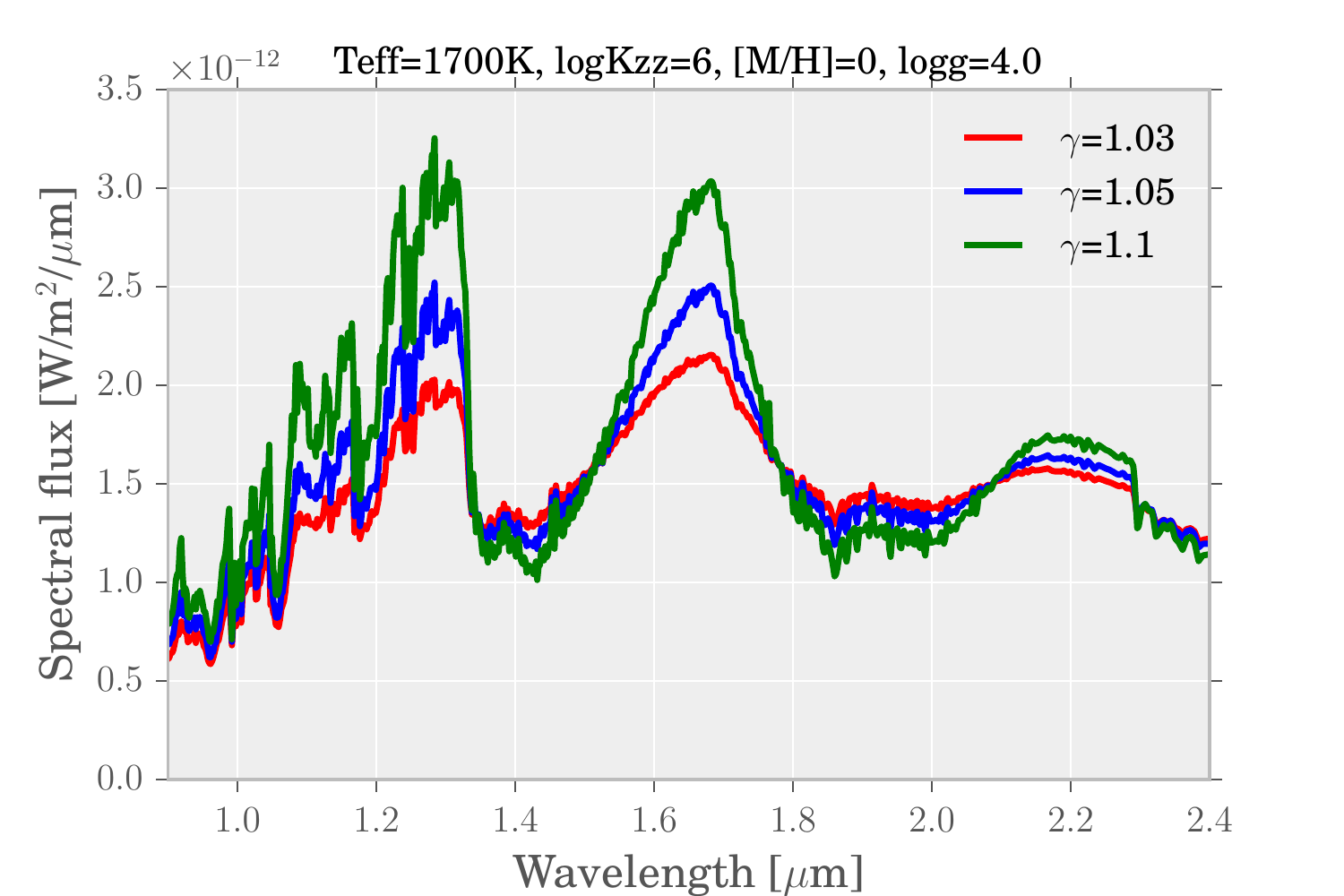}
\includegraphics[width=0.48\linewidth]{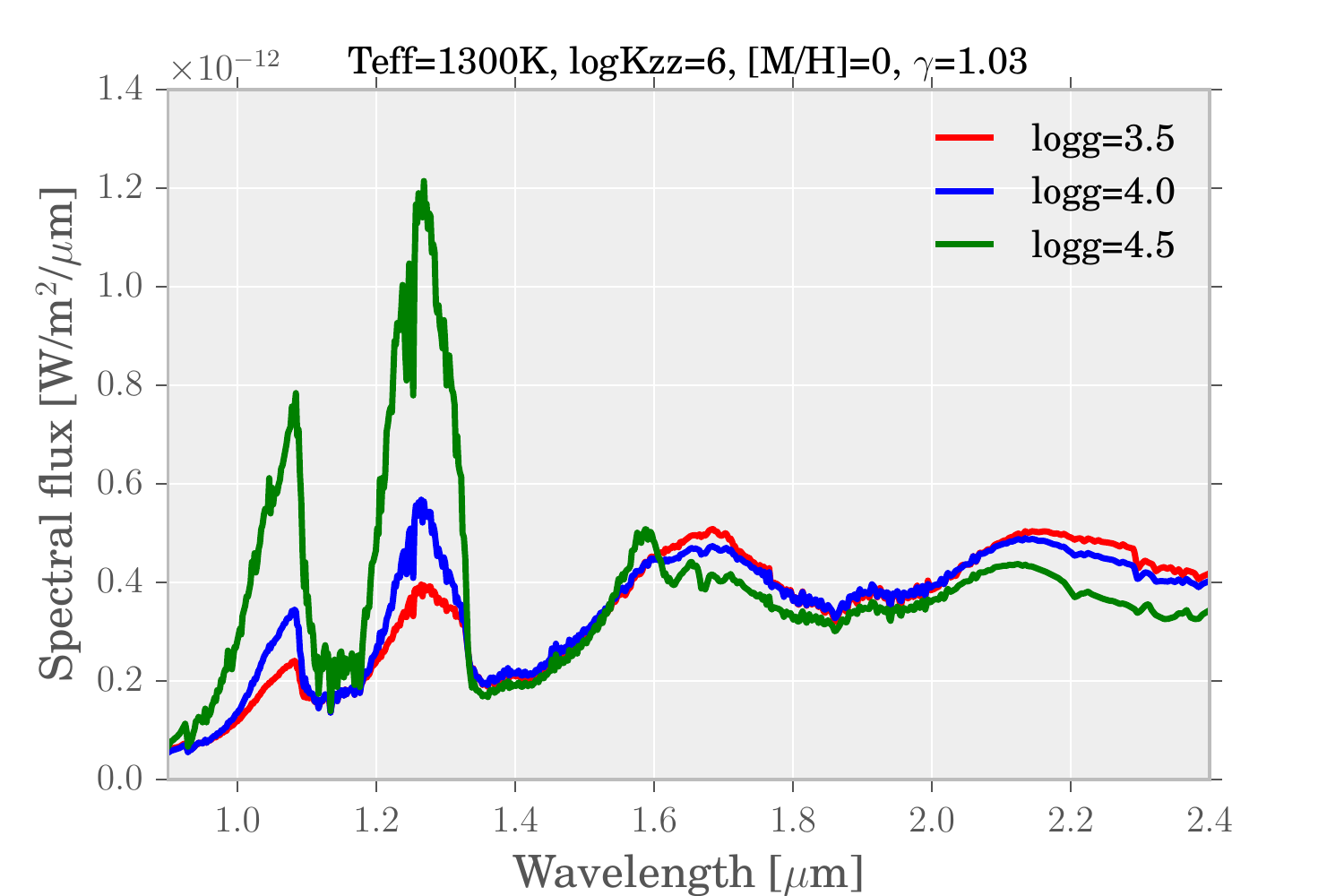}
\caption{\label{fig:models} Effects of varying different parameters on the spectral models obtained with the \texttt{ATMO} code.}
\end{figure*}

%%%%%%%%%%%%%%%%%%%%%%%%%%%%%%%%%%%%%%%%%%%%%%%%
%
% EGP/L dwarfs modeling
%
%%%%%%%%%%%%%%%%%%%%%%%%%%%%%%%%%%%%%%%%%%%%%%%%
\begin{figure*}[t]
\centering
\includegraphics[width=\linewidth]{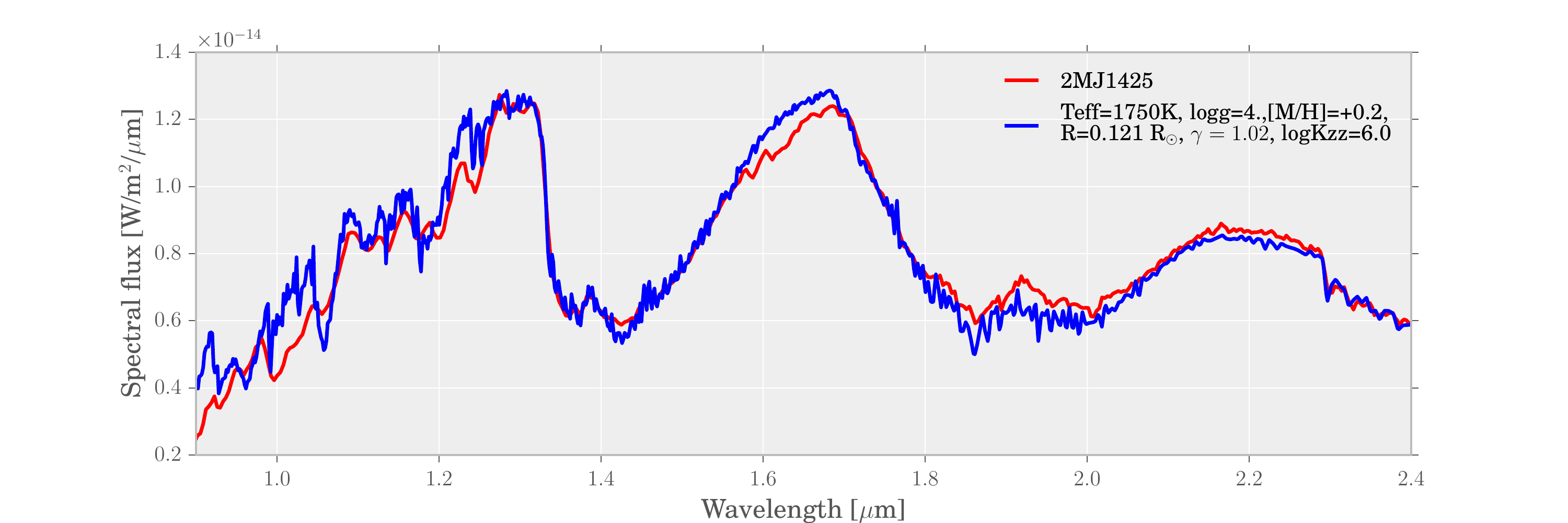}
\includegraphics[width=\linewidth]{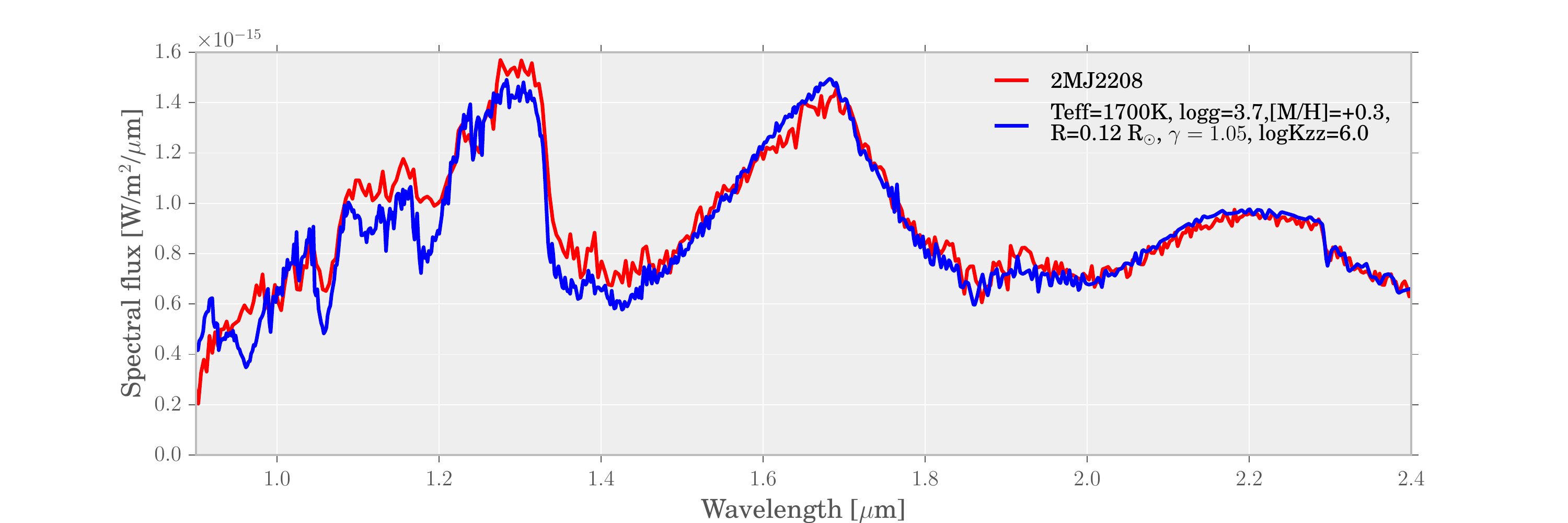}
\includegraphics[width=\linewidth]{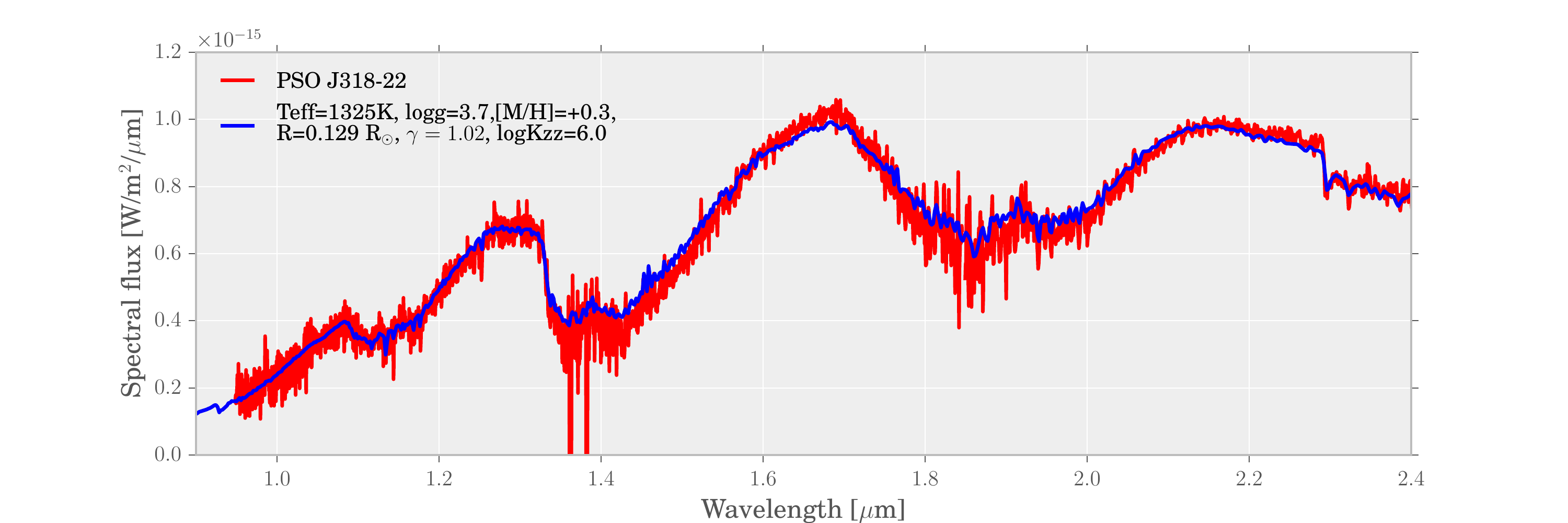}
\caption{\label{fig:obs}  Spectral models obtained with the \texttt{ATMO} code using a temperature gradient reduction in the atmosphere \citep{Tremblin:2016hi} compared with the SpeX prism spectra of DENIS J1425279-365023 \citep{BardalezGagliuffi:2014fl} and 2MASSW J2208136+292121  \citep{Allers:2013hk}, and the GNIRS spectrum of PSO J318.5338-22.8603 \citep{Liu:2013gy}.}
\end{figure*}

\section{Spectral models}\label{sect:spec}

The spectral models are all done with the \texttt{ATMO} code \citep{Amundsen:2014df,Tremblin:2015fx,drummond:2016aa}. We kept the same opacity sources as in \citet{Tremblin:2016hi} for direct comparison (H$_2$-H$_2$, H$_2$-He, H$_2$O, CO, CO$_2$, CH$_4$, NH$_3$, K, Na, TiO, VO, and FeH, , from the high temperature ExoMol \citep{Tennyson:2012ca} and HITEMP \citep{ROTHMAN20102139} line list databases when available). 

Similarly to \citet{Tremblin:2015fx,Tremblin:2016hi,drummond:2016aa}, we use the coupling of the radiative/convective code with the CHNO-based chemical network of \citet{Venot:2012aa} for the treatment of the departure from chemical equilibrium in the nitrogen and carbon chemistry. Starting from a converged pressure/temperature (PT) structure at chemical equilibrium, we integrate the time evolution of the abundances of all the chemical species in all the layers of the atmosphere. When turbulent mixing is faster than the chemical timescales (mixing coefficient K$_\mathrm{zz}$\footnote{K$_\mathrm{zz}$ is a free parameter and represents the vertical eddy diffusion coefficient}), the abundances of some molecules are not at chemical equilibrium anymore and can be controlled by the turbulent diffusion from the deepest layers. The non-equilibrium abundances are computed by solving the time evolution of the continuity equation \citep[see Eq. 10 in][]{drummond:2016aa} with the LSODE\footnote{https://computation.llnl.gov/casc/odepack/} integrator for stiff ordinary differential equations \citep{Hindmarsh:1983aa}. We re-converge the PT structure with the radiative/convective code regularly during the time integration until we reach a steady state for both the chemistry and energy conservation/hydrostatic equilibrium. We then get a coherent atmospheric structure with out-of-equilibrium chemical abundances. As in \citet{Tremblin:2016hi}, we propose that a process similar to fingering convection is responsible for this turbulent mixing and leads to the out-of-equilibrium abundances. Fingering convection can be triggered in the atmospheres of brown dwarfs and exoplanets because of the gradient of mean molecular weight induced by the chemical transitions CO/CH$_4$ and N$_2$/NH$_3$ at chemical equilibrium. Similarly to convective motions that lead to an adiabatic PT structure in the deep atmosphere, fingering convection could impact the PT structure of the atmosphere. As in \citet{Tremblin:2016hi}, we propose that fingering-convective motions can induce a temperature gradient reduction leading to a reddening in the modeled spectrum. This temperature gradient reduction in the atmosphere is modelled using an artificially reduced adiabatic index $\gamma$ in between two pressure levels on top of the convective zone.

In Fig.~\ref{fig:models}, we show the influence of various model parameters at an effective temperature of 1700~K and 1300~K, representative of the objects studied in this letter. At 1700~K we can see a degeneracy between gravity and metallicity on the spectral shape of the $H$ and $K$ bands. Decreasing gravity or increasing metallicity tends to form a triangular $H$ band and a strong and flatter $K$ band. Decreasing the temperature gradient in the atmosphere by decreasing the effective adiabatic index $\gamma$, increases the reddening in $J-K$. Increasing the surface gravity at an effective temperature of 1300~K has a strong effect on out-of-equilibrium chemistry and the quenching of CO in the deep atmosphere. At higher gravities, the CO/CH$_4$ transition is deeper in the atmosphere, hence it is more difficult to prevent the formation of CH$_4$ (strongly appearing at $\log g$=4.5) for a given mixing coefficient K$_\mathrm{zz}$. Thus varying K$_\mathrm{zz}$ would approximately lead to the same effect as varying gravity for T$_\mathrm{eff}$ = 1300~K. For the fitting of the observations we have probed a similar parameter space $\log g$ in [3.5, 4.5], [M/H] in [0, 0.5], T$_\mathrm{eff}$ in [1200, 1800], and $\gamma$ in [1.01, 1.1].   

\subsection{DENIS J1425279-365023}
DENIS J1425279-365023 was first identified as an L5 dwarf \citep{Kendall:2004kb} with a trigonometric distance of $\sim$~11.57$\pm$0.11~pc \citep{Dieterich:2014ic}. A SpeX Prism obtained by \citet{Gagne:2015dc} indicates that the object is a L4{\sc INT-G} object based on the index-based classification of \citet{Allers:2013hk}. Based on the galactic position and space velocity of the object, \citet{Gagne:2015ij,Gagne:2015dc} have also concluded that DENIS J1425-36 is a bona field member of the AB Doradus Moving Group \citep[149$^{+51}_{-19}$ Myrs,][]{Bell:2015aa}.

Figure~\ref{fig:obs} shows the SpeX prism spectrum of DENIS J1425-36 compared to a model obtained with the \texttt{ATMO} code. As in \citet{Tremblin:2016hi}, the reddening of the modeled spectrum is obtained by a temperature gradient reduction in the atmosphere using an artificially reduced adiabatic index $\gamma$. \citet{Tremblin:2016hi} have obtained a very good fit to field standard L dwarfs with a $\gamma$ value around 1.05 in the region between $\sim$~100~bars and $\sim$~2~bars. This result has recently been confirmed by \citet{BenBurningham:2017vq} although the authors suggest that sophisticated cloud models could provide a slightly better fit. For DENIS J1425-36, we show that the same model provides a good fit to the observed spectrum with a shallower temperature gradient with $\gamma\approx 1.02$ extending up to 0.25~bars in the atmosphere. As anticipated from the $H$ and $K$ band spectral shapes, we also find a low surface gravity, which indicates the youth of the object but also a slightly supersolar metallicity. Most importantly, we have obtained a radius of $\sim$~0.12 R$_\odot$ for DENIS J1425-36 which corresponds to an age of 100-150 Myrs, consistent with evolutionary models \citep{Baraffe:2003bj} and the age of the AB Doradus Moving Group. 

\subsection{2MASSW J2208136+292121}

2MASSW J2208136+292121 was first discovered by \citet{Kirkpatrick:2000gi} and classified as a peculiar L2. The peculiarity was later understood as signs of low gravity and the object is now classified as a L3{\sc VL-G} with a parallax distance of 47.6$\pm$2~pc \citep{Liu:2016co}. Based on its optical spectrum, \citet{Kirkpatrick:2008ec} estimated an age of $\sim$~100 Myr for the object. \citet{Gagne:2014gp} has estimated a probability of only 10~\% for a potential membership to the Beta Pictoris Moving Group, which would suggest a younger age of 24$\pm$3 Myrs \citep{Bell:2015aa}. 

Figure~\ref{fig:obs} shows the SpeX prism spectrum of 2MASS J2208+29 compared to a model obtained with the \texttt{ATMO} code. 
This object appears to have a CMD position discrepant with its gravity classification \citep{Liu:2016co} with a shift of only 0.1-0.2 mag in $J-K$ relative to the standard field objects. This fact directly reflects in the effective $\gamma$ value of 1.05 found with the spectral model, a value closer to the one used for standard field object \citep{Tremblin:2016hi}. The shape of the spectrum, e.g. the very peaked $H$ band, clearly indicates a low gravity and/or a high metallicity for the object, as found with our model. We have found a radius similar to the one of DENIS J1425-36, around 0.12 R$_\odot$. This radius corresponds to an age of 100-150 Myrs with evolutionnary models. This result confirms the ``optical'' age found by \citet{Kirkpatrick:2008ec} and argue against a membership to the Beta Pictoris Moving Group. 

\subsection{PSO J318.5338-22.8603}

PSO J318.5338-22.8603 has been identified as the reddest field object ever observed \citep{Liu:2013gy} with a parallax distance of 24.6$\pm$1.1~pc. Its $J-K$ color is 2.78 mag, which has been interpreted as the sign of an unusually dusty atmosphere. This interpretation, however, is weakened by the difficulty of fitting the NIR spectrum with cloud models. A good fit can be obtained with BT-Settl models \citep{Liu:2013gy} but with an implausibly small radius (around 0.08 R$_\odot$) for such a young object. Its membership to the Beta Pictoris Moving Group suggests an age around 24$\pm$3 Myrs  \citep{Bell:2015aa}.

Figure~\ref{fig:obs} shows the GNIRS\footnote{ Gemini NIR Spectrometer \citep{elias:2006aa}.} spectrum of PSO J318-22 compared to a model obtained with the \texttt{ATMO} code. The correspondence between the model and the observed spectrum is remarkably good and emphasizes the high precision of the H$_2$O and CO high-temperature opacities derived from ExoMol \citep{barber:2006aa} and HITEMP linelists. Similarly to DENIS J1425-36 and 2MASSJ2208+29, the spectral shape of $H$ and $K$ bands points toward a low-gravity and/or high-metallicity atmosphere which is indeed indicated by the modeled spectrum. The reddening is obtained with an effective $\gamma$ of 1.02 up to 0.1 bars in the atmosphere. Out-of-equilibrium chemistry of CO/CH$_4$ plays an important role in the atmosphere of PSO J318-22. Unlike DENIS J1425-36 and 2MASSJ2208+29, that have too high temperatures, PSO J318-22 is sufficiently cold for CH$_4$ to dominate the atmosphere. Out-of-equilibrium processes \citep[that can be linked to fingering convection][]{Tremblin:2016hi} are crucial to prevent CH$_4$ formation and keep carbon in CO as expected from the observed spectrum, due to quenching of CO from the deep atmosphere where it is favoured in chemical equilibrium. In our models, PSO J318-22 is actually right at the transition of having CH$_4$ as the dominant carbon-bearing species. The radius we have obtained is around $\sim$0.13 R$_\odot$ which corresponds to an age between 30-50 Myrs, in a good agreement with the estimated age of the Beta Pictoris Moving Group.

\begin{figure*}[t]
\centering
\includegraphics[width=\linewidth]{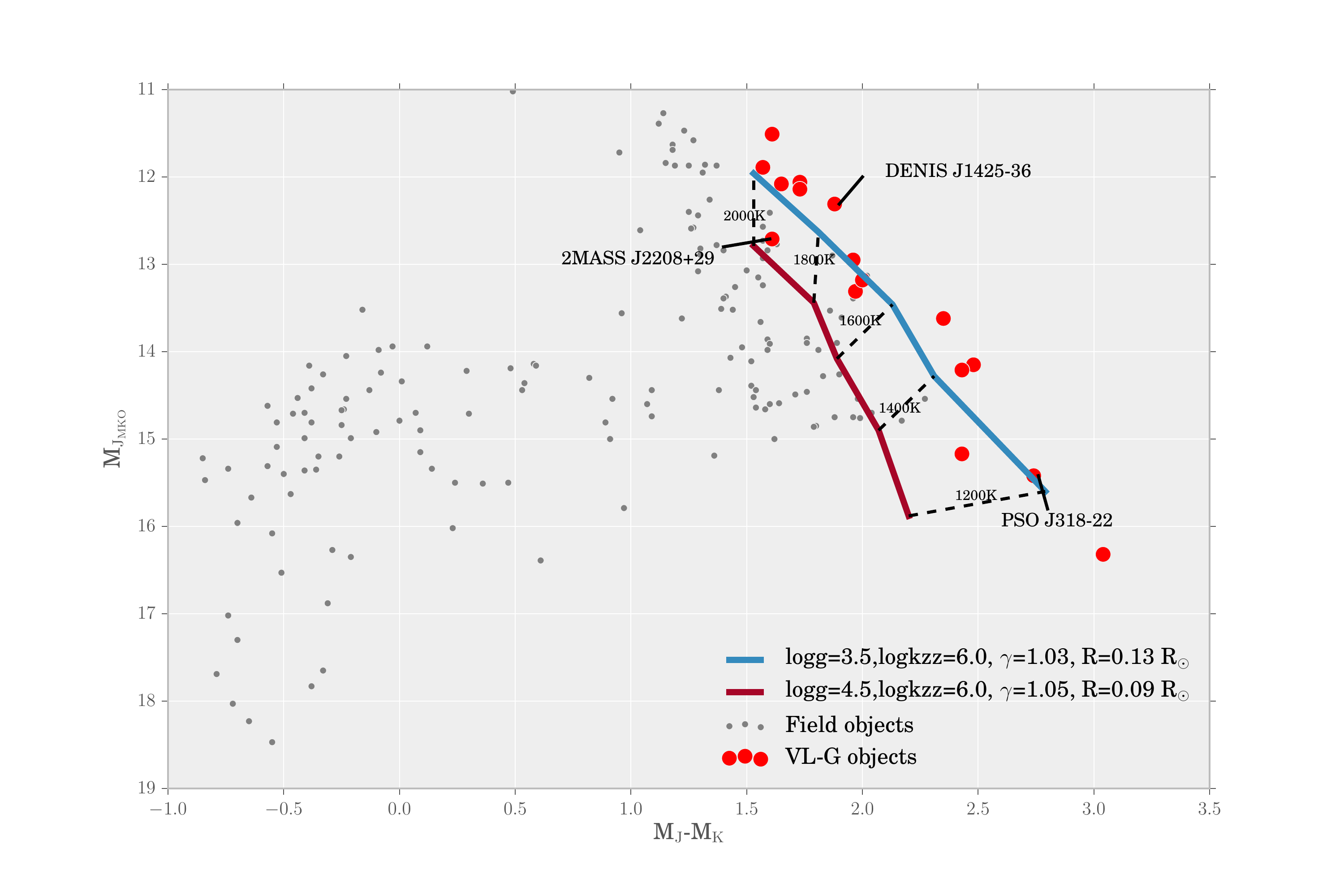}
\caption{\label{fig:cmd} M$_J$ versus $J-K$ MKO color magnitude diagram of field objects \citep{Dupuy:2012bp,Faherty:2012cy}, VL-G objects \citep{Liu:2016co}, and the \texttt{ATMO} model sequence with varying effective temperature.}
\end{figure*}

\begin{figure}[!t]
\centering
\includegraphics[width=\linewidth]{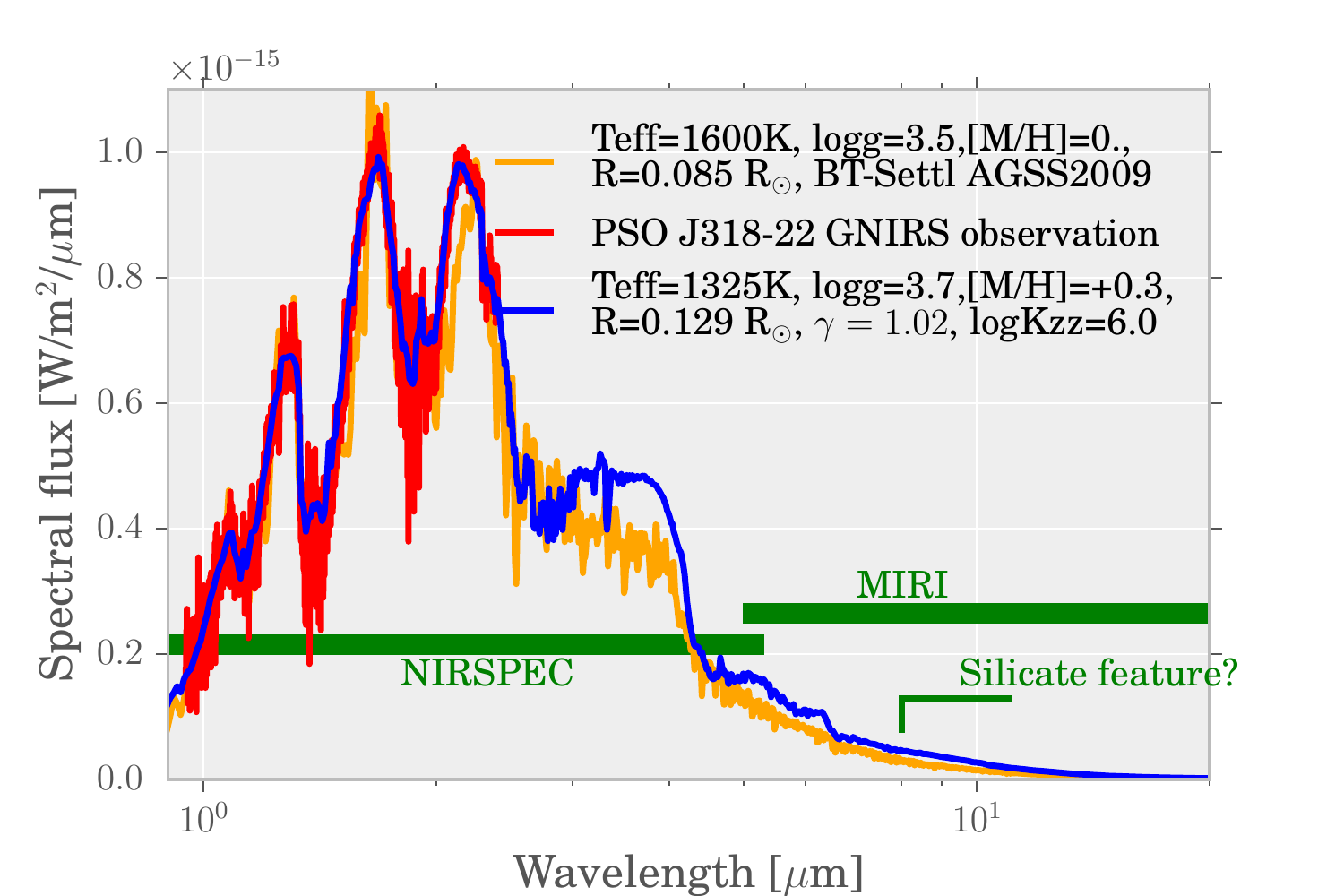}
\includegraphics[width=\linewidth]{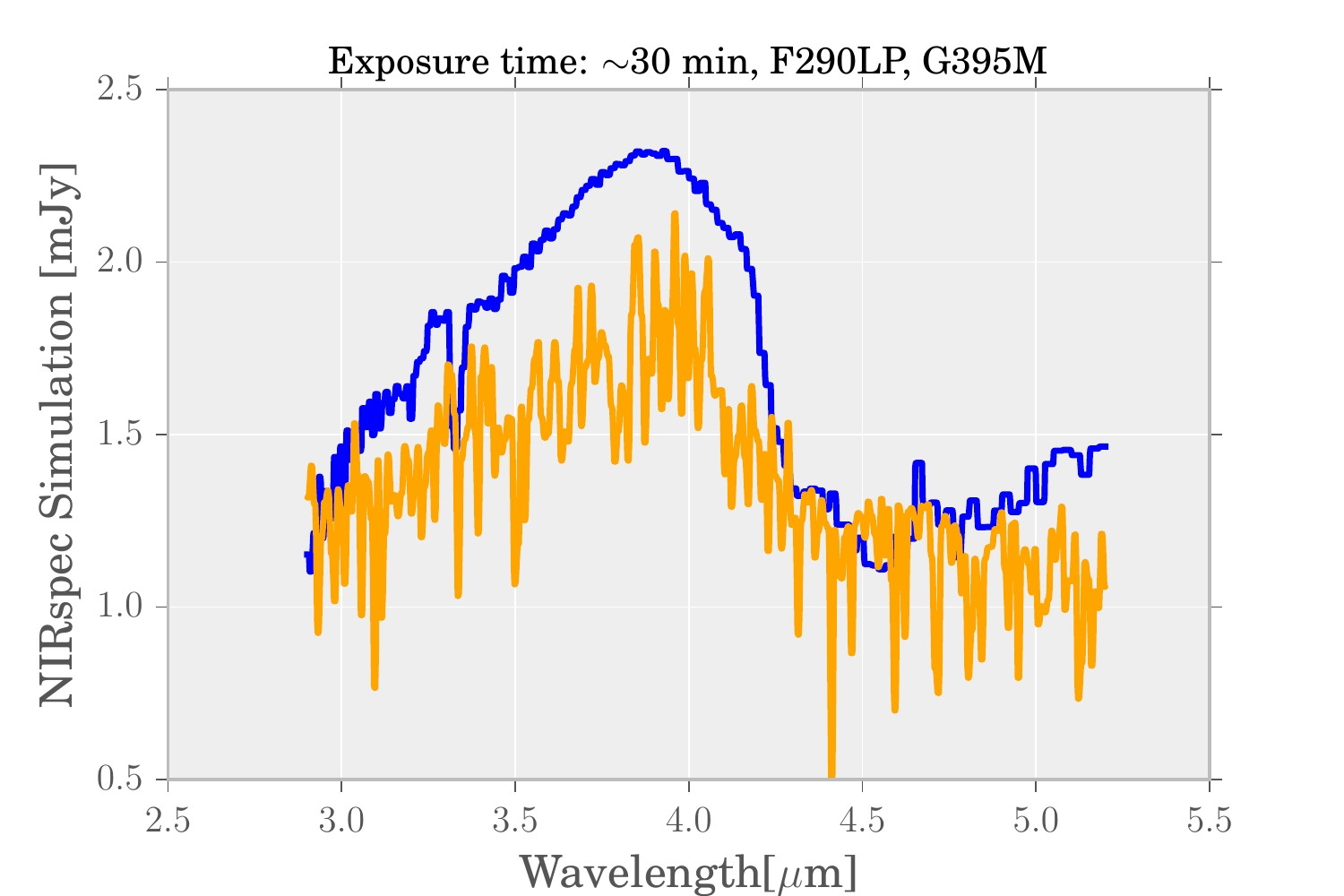}
\caption{\label{fig:pso_jwst} Top: full spectral model of PSO J318-22 with NIRspec/MIRI spectral coverage. Bottom: NIRspec simulation using the grating G395M with a $\sim$30 min exposure time.}
\end{figure}

\subsection{Color-Magnitude diagram}\label{sect:cmd}

Figure~\ref{fig:cmd} shows the CMD M$_J$ versus $J-K$ for field objects and VL-G objects \citep{Liu:2013gy}. The sequence of models has been computed at a fixed surface gravity of $\log g$=3.5, radius of 0.13~R$_\odot$, vertical mixing at 10$^{6}$~cm$^2$/s, effective adiabatic index of $\gamma$=1.03, and with solar metallicity. The range of effective temperature is between 1200~K and 2000~K, which roughly corresponds to objects of masses between 5 and 20 M$_\mathrm{jup}$. As shown in Fig.~\ref{fig:models}, the effect of gravity and metallicity will be relatively small on the $J$ and $K$ band magnitude, compared to the effect of varying $\gamma$. As expected from the spectral models, this sequence reproduces well the reddening of VL-G objects in the CMD. A sequence with a higher $\gamma$, e.g. 1.05 as for 2MASS J2208+29, would reproduce the field objects with a smaller reddening. Cloud models are currently struggling to reach the reddening of the VL-G objects. \citet{Liu:2016co} found that the models from \citet{Saumon:2008im} and \citet{Marley:2012fo} are systematically too blue. BT-Settl models \citep{allard:2014aa} tend to also be bluer than the VL-G objects, and faint red L dwarfs are not reproduced because the cloud clearing occurs at brighter magnitudes.  It is worth stressing that, with the cloud interpretation, these latter models show that there is no correlation between reddening and the CO/CH$_4$ transition, although observations do show such a correlation both for standard field objects and for VL-G objects. 

\section{Discussion and JWST perspectives}\label{sect:jwst}

For DENIS J1425-36 and 2MASS J2208+29, the correspondence between the modeled and observed spectra is better in the 1.4-2.4 $\mu$m region than below 1.4 $\mu$m. This limitation could be caused by our opacity database that currently contains TiO, VO, and FeH but may lack other metal hydrides that could be important at high effective temperatures around 1700~K. As seen in Sect.~\ref{sect:spec}, the spectral shape of the $H$ and $K$ bands clearly indicates a low gravity and/or high metallicity, a degeneracy already observed for other objects \citep[e.g.][]{Delorme:2017uc}. Dedicated studies of the metallicity of moving groups point towards low values, e.g. [M/H]=0.1 for AB Doradus \citep{Biazzo:2012ht}. Our models are subject to the same degeneracy and we have generally chosen to keep a metallicity around [M/H]=0.2-0.3 to be consistent with the metal Galactic distribution \citep{Anders:2017ii}. For these slightly higher metallicities, we still obtain surface gravities $\sim$0.5 dex lower than expected from evolutionary models \citep{Baraffe:2003bj}. Since these evolutionary models are at solar metallicity, the computation of coherent models at higher metallicity might resolve this discrepancy. Future studies using retrieval analysis will also be useful to better quantify the degeneracy and help assessing if this bias towards high metallicity/low gravity is indeed an issue. 

Despite the slightly too-low gravities predicted by our models, the derived radii of the objects are in good agreement with evolutionary models and the ages of the moving groups. Although clouds are predicted to be thicker, and hence easier to detect/characterize in the cloud interpretation, cloudy models have struggled to produce reasonable radii, e.g. in the case of the low-gravity very red HR 8799 planets \citep[see summary in][]{Marley:2012fo}. Therefore, our VL-G models strengthen the interpretation of \citet{Tremblin:2016hi} that NIR reddening is not necessarily a consequence of the presence of clouds near the photosphere, but could be the signature of fingering convection induced by the CO/CH$_4$ chemical transition.

By using a simple model, we can estimate the magnitude of the velocities that need to be triggered by fingering convection in order to obtain an effective adiabatic index around 1.03. Following \citet{Tremblin:2017aa}, we can write the steady state energy conservation equation in the following form:
\begin{eqnarray}
  u_{r,\mathrm{conv}}\left(\nabla_T-\frac{\gamma_\mathrm{ad}-1}{\gamma_\mathrm{ad}}\right) &=& \frac{\gamma_\mathrm{ad}-1}{\gamma_\mathrm{ad}}\frac{H_\mathrm{rad}}{\rho g}\cr
  u_{r,\mathrm{conv}}\left(\nabla_T-\frac{\gamma_\mathrm{eff}-1}{\gamma_\mathrm{eff}}\right) &=& 0 \cr
  \frac{\gamma_\mathrm{eff}-1}{\gamma_\mathrm{eff}}&=&  \frac{\gamma_\mathrm{ad}-1}{\gamma_\mathrm{ad}}\left(1+\frac{H_\mathrm{rad}}{u_{r,\mathrm{conv}} \rho g} \right)
\end{eqnarray}
With $\nabla_T = \partial \ln T/\partial \ln P$, $ u_{r,\mathrm{conv}}$ the vertical (overturning or fingering) convective velocities, $H_\mathrm{rad}$ the radiative heating rate, $\rho$ the density, $g$ the gravity, $\gamma_\mathrm{ad}$ the standard adiabatic index of the gas, and $\gamma_\mathrm{eff}$ the effective adiabatic index defined by the last equation. In the overturning convective part of the atmosphere, $H_\mathrm{rad}\approx 0$ and the atmosphere is adiabatic with $\nabla_T\approx (\gamma_\mathrm{ad}-1)/\gamma_\mathrm{ad}$. In the fingering convective part, the atmosphere is in a zone where the radiative energy transport cannot be ignore, hence  $\nabla_T\approx (\gamma_\mathrm{eff}-1)/\gamma_\mathrm{eff}$ with $\gamma_\mathrm{eff} \neq \gamma_\mathrm{ad}$. We can give an order of magnitude estimation of the velocities that are needed to obtain $\gamma_\mathrm{eff}\approx 1.03$. For the model of PSOJ318-22, we estimate that fingering-convective velocity of the order of 10 m/s are required to obtain such a value of the effective adiabatic index (for comparison velocities in the overturning convection zone are in the range 30-100 m/s). 

As in \citet{Tremblin:2016hi}, we emphasize that this does not mean that clouds are not present in the atmosphere of these objects. Thin high-altitude clouds are likely to form and be driven to the top of the fingering-convective zone in the atmosphere. They could be observed e.g. in the silicate absorption band at 10 $\mu$m \citep[][]{Cushing:2006hx}. Although we need to expand the fingering convective zone up to $\sim$~0.1~bar in the atmosphere, it is not clear whether this is sufficient to observe thin high-altitude clouds since the models reach a 10-$\mu$m optical depth of 1 at $\sim$~0.01 bars. Key future constrains on these signatures will be obtained by JWST with MIRI (see the indicative location of the absorption feature in the MIRI spectral coverage in Fig.~\ref{fig:pso_jwst}). 

As a conclusion, we recall that fingering convection induced by the CO/CH$_4$ chemical transition can naturally explain the nature of the L/T transition, a transition between fingering-convective ``red'' CO-dominated atmospheres to stable `''blue'' CH$_4$-dominated ones. The reddening of the spectrum of field and VL-G objects can be very well explained by a temperature gradient reduction caused by fingering convection. The estimated radii from the models are in good agreement with evolutionary models and the estimated age of moving groups, which strengthens the fingering interpretation.
%% \textbf{The interpretation of the L/T transition with cloud models could be the formation of holes leading to the dissipation of the cloud layer \citep{Burrows:2006aa}, however, recent works \citet{Buenzli:2015aa} seem to rule out the existence of such holes and would point towards cloud height variations. It is then difficult to make the clouds disappear quickly as a function of effective temperature, and the physical mechanisms behind holes/height variations remain relatively mysterious. }

Nevertheless, as shown in \citet{BenBurningham:2017vq}, both cloud and cloudless models can provide a very good fit to the spectra of field L dwarfs, and cloud models have generally shown good successes in the spectral modelling of field L and T dwarfs in the past 20 years. Young low-gravity brown dwarfs could provide a good test to distinguish between the two models if the current limitations of cloudy VL-G models are confirmed by future studies.
We provide in Fig.~\ref{fig:pso_jwst} the long wavelength (3-20 $\mu$m) prediction of our full spectral model for PSO J318-22 compared with a BT-Settl cloudy model and simulated observations with NIRspec using the grating G395M with $\sim$30 min exposure time. JWST with NIRspec and MIRI observations (SNR $\gtrsim$ 100) in the 3-7 $\mu$m region and future comparisons with cloud models in that wavelength range could help assess which interpretation between clouds and fingering convection is correct. 

%%%%%%%%%%%%%%%%%%%%%%%%%%%%%%%%%%%%%%%%%%%%%%%%
%
% Acknowledgments
%
%%%%%%%%%%%%%%%%%%%%%%%%%%%%%%%%%%%%%%%%%%%%%%%%

\begin{acknowledgements}
This work is partly supported by the European Research Council under the European Community's Seventh Framework Programme (FP7/2007-2013 Grant Agreement No. 247060-PEPS and grant No. 320478-TOFU). MCL acknowledges support from NSF grant AST-1518339. POL acknowledges support from the LabEx P2IO, the French ANR contract 05-BLAN-NT09-573739.
\end{acknowledgements}

%%%%%%%%%%%%%%%%%%%%%%%%%%%%%%%%%%%%%%%%%%%%%%%%
%
% Bibliography
%
%%%%%%%%%%%%%%%%%%%%%%%%%%%%%%%%%%%%%%%%%%%%%%%%

%\appendix

%% \bibliographystyle{abbrvnat}
%%  \bibliography{main.bib}

\end {document}